# Binary Contact Sensing for Sitting Posture Recognition Without Pressure Sensors

Orthy Toor[1], Khandaker Mashiat Rahman[1], Tonoya Mustafa[1], Abdullah Bin Shams[2*]
[1]Dept. of Electrical and Electronic Engineering, Islamic University of Technology, Gazipur 1704, Bangladesh
[2]Dept. of Electrical & Computer Engineering, University of Toronto, Toronto, Ontario M5S 3G4, Canada
*Corresponding author: ab.shams@utoronto.ca

***Abstract*—Prolonged sitting with poor posture results in musculoskeletal injury. Early intervention and prevention methods to monitor posture rely on cameras, wearable devices, and dense pressure arrays. Although effective, these approaches introduce privacy concerns, calibration needs, higher cost, etc. In this paper, we explore the spatial pattern of body contact as a binary posture feature vector for a distinct contact-based sitting posture recognition system without the need for analog signal conditioning and calibration. Our sensing principle uses 10 mechanical contact switches arranged in a 5 x 2 array on the backrest. Each switch encodes local body contact into 10-bit binary posture signatures for four postures: normal sitting, leaning back, leaning left, and leaning right. Decision tree and logistic regression classifiers achieved highest accuracy of 96%. Low correlation amongst the switches indicates that every switch captures complimentary information for successful posture classification. SHAP analysis identified the central contacts as the most informative and significant region for posture discrimination. Our results show that binary contact patterns can capture sufficient spatial information for reliable posture recognition. This sensing strategy offers a simple & low-cost alternative to pressure-based systems to monitor posture without the need for mapping the biomechanical pressure distribution.**

## I. Introduction

Sitting posture is critical to maintain a healthy spine. It determines how stress is distributed amongst muscles and joints. At recent times, technological advancements in the form of computers and smartphones led to a rise in sedentary behavior, with people spending on average about 82% of their working hours sitting [1]. Sitting in the wrong posture for an extended period increases pressure from neck-&-shoulders to lower back, inducing muscle fatigue, back pain, neck pain, poor body posture, and other musculoskeletal conditions. This affects around 1.71 billion people globally, with 570 million people suffering from low back pain, establishing musculoskeletal disorder the leading cause of disability in 160 countries [2]. In response to this escalating health crisis, researchers are prioritizing developing posture monitoring and correction systems.

Researchers have explored various techniques to identify posture. Camera with computer vision algorithms have been used to develop real time multiple individual posture monitoring system [3]. These camera-based systems highly depend on camera position, camera resolution, proper lighting, and a clear view of the subject. As the subjects are continuously recorded, there is also a privacy concern issue. Wearable devices with inertial sensors have been integrated with cloud communication, and mobile application for continuous posture monitoring [4]. However, subjects must wear the device throughout the monitoring period. Wi-Fi Channel State Information (CSI) based device-free wireless sensing techniques have also been explored for human activity recognition by analyzing various wireless signals without requiring deployment of cameras or wearable devices solving the privacy issue. However, it depends highly on how the environment influences the signals [5]. Also, the sensors need correct positioning and calibration, which may reduce subjects comfort during daily use. Pressure-based sensing is commonly used for sitting posture detection, using Arduino based interface with Force Sensitive Sensors (FSR) for real-time posture monitoring, posture asymmetry detection and clou-based monitoring [6]–[9]. However, FSR-based systems are designed mainly to quantify mechanical load distribution rather than directly determining sitting posture. Since posture classification mainly depends on identifying the spatial arrangement of contact between the subject and the chair, continuous pressure measurements provide data that may not require for posture classification. Moreover, pressure distribution is highly dependent on subject-specific factors, including body weight, physical dimensions, clothing, and preferred sitting style, even when the same posture is maintained. These variations make pressure-based posture recognition more complex, often requiring additional signal processing, calibration, force normalization, and feature extraction to achieve reliable posture classification.

To circumvent these limitations, this work proposes a binary contact (two-state) based sensing mechanism that directly monitors body-chair interaction. Mechanical contact switches are positioned on the chair to detect two discrete sates: contact or no contact. The resulting binary contact patterns offer a simple representation of the spatial characteristics of different sitting postures without capturing force-related information. As the sensing mechanism is based on physical contact rather than pressure magnitude, the system is less sensitive to variations in subject characteristics and gradual changes in applied force over prolonged sitting periods. In addition, the digital output generated by the contact switches simplifies the sensing mechanism by eliminating analog signal conditioning, continuous calibration, and complex feature extraction, allowing an alternate efficient real-time implementation of embedded microcontroller-based platforms for posture classification.

## II. Literature Review

Existing studies on sitting posture recognition have utilized various sensing technologies, such as vision-based systems, wearable sensors, pressure and force sensors, and smart chair systems. Among the various sensing techniques, Vision-based methods have shown strong performance in posture recognition. An OpenVINO and artificial intelligence-based framework demonstrated real-time activity recognition capabilities [3], an F1-score of 98.1% for Sit Pose based on an Azure Kinect depth camera and ensemble learning [1]. Moreover, deep learning models, including InceptionV3-SVM [10] and MobileNetV2-LSTM [11] , achieved recognition accuracies greater than 99%. However, these methods often require suitable camera positioning and lighting conditions. In addition, their computational complexity, occlusion-related limitations, and privacy-related concerns may reduce their suitability for continuous posture monitoring in real-world environments.

TABLE I
Comparison of Existing Posture Recognition Systems

| Ref. | Sensor Type | Sensor Position | Accuracy (%) |
|---|---|---|---|
| [3] | RGB Camera | Front View | 95.9 |
| [4] | IMU Smart Belt | Waist | – |
| [7] | Pressure Sensors | Chair Seat | 99.10 |
| [8] | Pressure Sensors | Chair Backrest | 98.82 |
| [9] | Pressure Mat | Seat & Backrest | 86 |
| [12] | 9-Axis IMU + LoRa | Wearable (Upper Body) | 95 |
| [13] | 6 FSR Sensors | Chair Seat and Backrest | – |
| [14] | 4 FSR Sensors | Seat Cushion | – |
| **Proposed Work** | **Binary Contact Switches** | **Chair Backrest** | **96.08** |

Wearable devices provide an alternative approach for posture monitoring by directly capturing body movements. An IMU-based smart belt was developed to support continuous posture tracking and provided real-time feedback through a mobile application [4]. In another study, a nine-axis IMU-based IoT system integrated with LoRa communication and a Random Forest classifier achieved an accuracy of approximately 95% [12]. However, the continuous attachment of wearable sensors may lead to subject discomfort and limit long-term usability. In addition, sensor calibration, wireless data transmission, and feature extraction can increase the complexity of system implementation.

Pressure and force-based sensing techniques have been widely adopted for posture recognition because of their ability to directly capture physical interactions between the subject and the chair. Previous studies have explored a variety of pressure and force sensing configurations, including force sensors [13], force-sensitive resistors (FSRs) [14], optimized pressure arrays [7], and full-back pressure sensing systems [8]. In addition, reduced-density pressure sensor layouts achieved approximately 85% accuracy in automotive applications [9], and textile-based pressure sensing achieved 97.9% accuracy in smart bed applications [15]. However, these approaches often depend on multiple processing stages, including analog signal conditioning, sensor calibration, force normalization, feature extraction, and machine learning algorithms. Their performance may also be affected by differences in subjects' body weight, body shape, and sitting habits.

Recent studies have increasingly focused on smart chair platforms that combine pressure, wearable, and vision-based sensors with artificial intelligence [6]. However, their practical implementation remains challenging due to high hardware costs, limited datasets, and high computational requirements. Conformal microwave sensing systems have achieved accuracies up to 99.7% using Random Forest and XGBoost [16]. However, their complex hardware makes them less suitable for widespread use.

These observations reveal a research gap for posture recognition systems that can provide reliable classification using simple sensing hardware and minimal signal processing. To address this gap, the proposed work uses a contact-switch-based approach. By monitoring binary contact on the chair, the system identifies the essential spatial contact patterns associated with predefined sitting postures without measuring the applied pressure. Since the system does not measure pressure magnitude, it avoids the need for pressure calibration, force normalization, and complex feature extraction. This makes the proposed system low-cost, robust, and suitable for real-time posture recognition.

## III. Methodology

### A. Hardware Design and Sensor Placement

The contact sensor of the system is built using ten binary contact switches which are push-button type. It turns on when the person's back presses the switch and gives a binary reading, ON or OFF. The binary switching pattern is recorded using Arduino UNO microcontroller. The layout of the switches is a 5×2 grid matrix where five rows of switches are positioned vertically, and two switch columns are positioned horizontally. The mentioned contact sensor is mounted on a standard rectangular veroboard. The veroboard is then secured on the backrest of a standard office chair. The measurement of backrest is 47 cm wide and 47 cm deep. The veroboard measures 15 cm by 6.5 cm, and it is secured on the backrest, with the 6.5 cm edge running along the chair's width and the 15 cm edge along its depth, which puts equal margins on all sides: 7.5 cm left and right, 4 cm top and 8 cm bottom. The sensor placement remains constant throughout the data collection. The ten binary values from the 10 switches are then encoded into a 10-bit row vector.

### B. Data Acquisiton

Data was collected from 12 subjects (6 male, 6 female), aged 21–24 years, with an average height of 1.65 m, weight of 70.4 kg, and body mass index (BMI) of 25.6 kg/m$^2$. Each subject was asked to sit on the instrumented chair and perform four postures: Normal Sitting, Leaning Back,

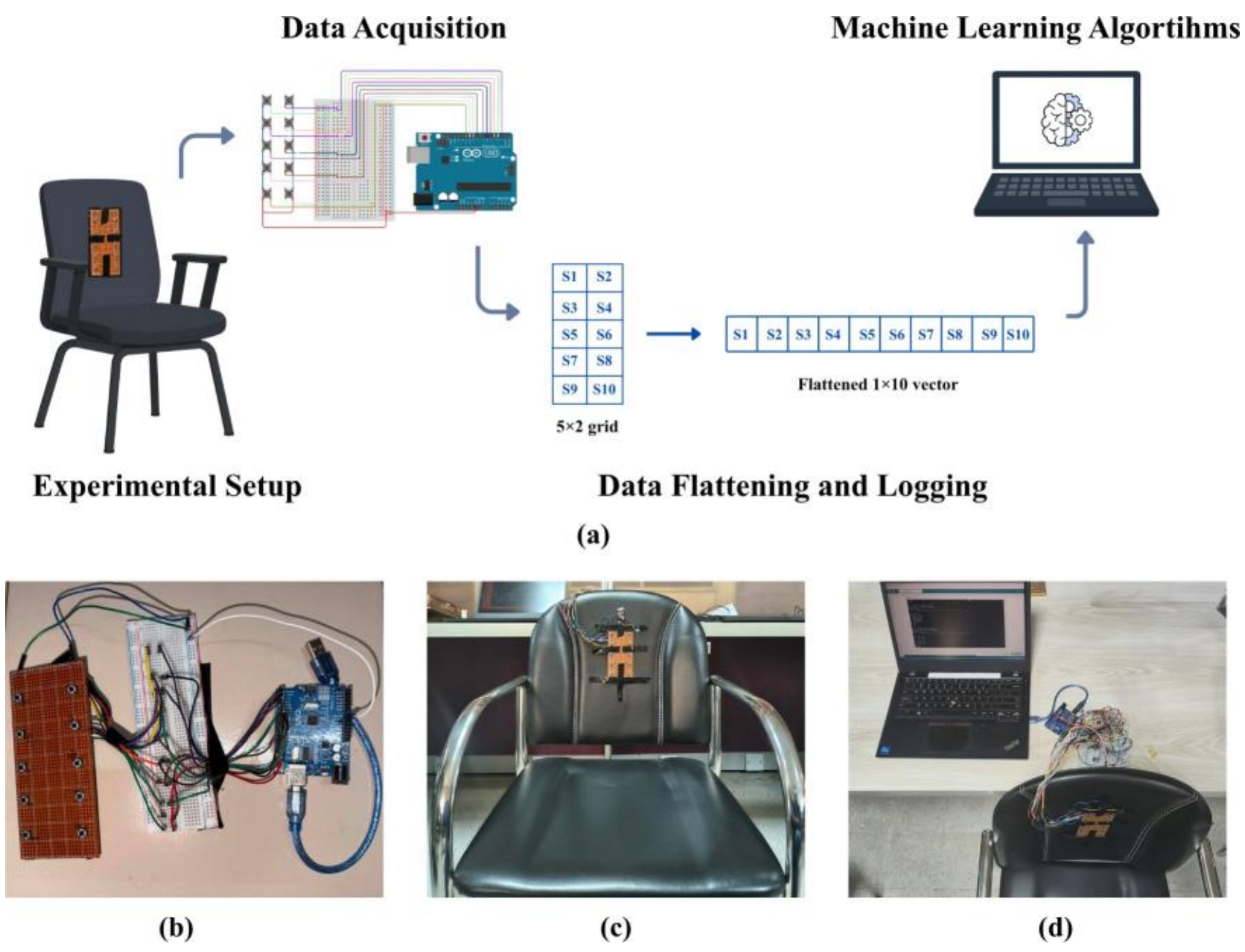

Fig. 1. (a) Outline of the experimental procedure (b) Circuit used in the study (c) Sensor board placement on chair (d) Experimental setup

Leaning Left, and Leaning Right. Normal sitting is similar to sitting upright on a chair, with the back straight. People sit differently for the leaning back posture. For different people sitting in a leaning-back posture, the upper and lower angles can have different combinations. Fig 2 demonstrate how we standardized the upper and lower angles for each participant, ensuring a standardized leaning-back posture. The upper lean angle was fixed at approximately 25$^{\circ}$ and lower lean angle was fixed at around 30$^{\circ}$. And the contact point was fixed at the middle of the sensor board.

For the leaning left and right postures, participants normally moved their torso left and right. Every recording session began with the posture label set in the microcontroller code first, and only then was the subject told to get into that position and hold it steady for a short stretch. Data recording started once the subject was properly positioned, so each sample gets correctly labeled at the source instead of after the fact. We generated a balanced dataset across the classes. In total we took 496 samples with 124 samples per class.

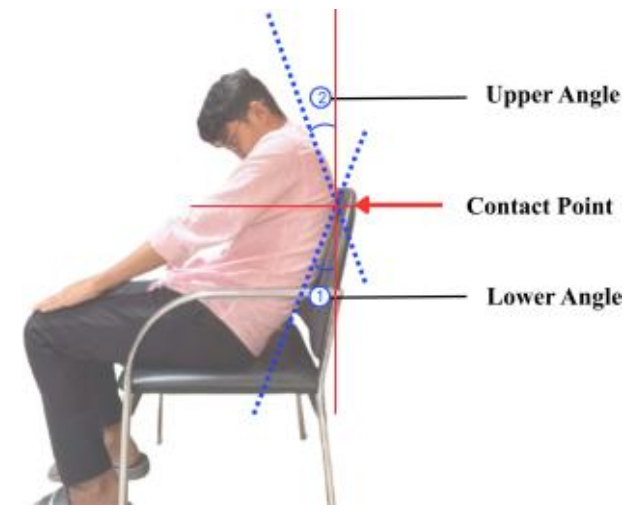

Fig. 2. Upper and lower lean angle standardization

### C. Signal-to-Feature Encoding

Depending on the subject's back contact with the set of 10 switches, they are either ON (1) or OFF(0). This produces a 5×2 switch matrix of ten binary readings, as demonstrated in Fig. 3. This matrix then gets converted into a 10-bit row vector, S1 through S10. We used the 10-bit vector directly as our one data sample, as our contact sensor's output is clean digital signals with no baseline drift or high-frequency noise to deal with, and each bit already relates to a specific spot on the chair. This eliminates filtering and feature extraction techniques which makes the whole system simple, fast, and practical for use on low-cost hardware.

### D. Train-Test Splitting and Cross-Validation

The complete balanced dataset was divided into an 80:20 ratio for training and testing sets, where 80% of the samples were used for training the models while the remaining 20% was used for validating the prediction model. [17], [18]. To avoid biases during training, a 10-fold stratified cross-validation was performed by dividing the training data into 10 equal folds. The model uses 9 set to train itself and tests on the one remaining, and this repeats 10 times where every fold gets its turn as the test set, which makes the training more robust. The results of all 10 folds were then averaged for reliable results.

### E. Machine Learning Algorithms and Evaluation Metrics

Six different machine learning classifiers were trained and compared for classifying the 4 different postures using the 10-bit binary row vector pattern from the contact sensors. The algorithms are: Logistic Regression (LR), K-Nearest Neighbors (KNN), Support Vector Machine (SVM), Decision Tree (DT), Random Forest (RF), and Extreme Gradient Boosting (XGBoost). Feature scaling was performed for LR, KNN, and SVM [19], [20]. Four evaluation metrics were used to evaluate and compare the six classifiers: accuracy, precision, recall, and F1-score.

### *F. SHAP and Switch Activation Frequency*

SHAP (SHapley Additive exPlanations) was performed to identify the contribution of each switch in the posture classification task. To further validate their respective contribution in actual physical contact pattern, the switch activation frequency was also analyzed.

## IV. Results and Discussion

The experimental setup consisted of a chair on which a board containing the 10 switches were mounted. The chair backrest had dimensions of 47 × 47 cm, where the switch array (15 × 6.5 cm) was mounted to maximize contact with the thoracic and lumbar regions while sitting. When a participant rests on the chair in different postures, their contact with the switch array generates binary patterns depending upon different postures. The ON/OFF states of the switches were acquired using an Arduino UNO and transmitted to a personal computer for real-time data logging and subsequent analysis. Twelve individuals voluntarily participated in data collection process. Four different postures were considered in the experiment: Normal sitting, Leaning Back, Leaning Left, and Leaning Right. One set of each posture was regarded as one sample, and in this way, 124 samples were collected from the 12 participants. The demographic characteristics of the participants, including average age, height, weight, and Body Mass Index (BMI), are summarized in III. The switch array is a 5×2 matrix, and when the ON/OFF pattern for each posture was collected, the 5×2 array was flattened into a 1×10 row vector to ease the classification process. The 1×10 switch positions were used as the input features of the Machine Learning models, with four postures being the four different classes. To evaluate the effectiveness of the proposed sensing approach, six supervised machine learning classifiers were investigated: RF, SVM, KNN, XGBoost, DT, and LR.

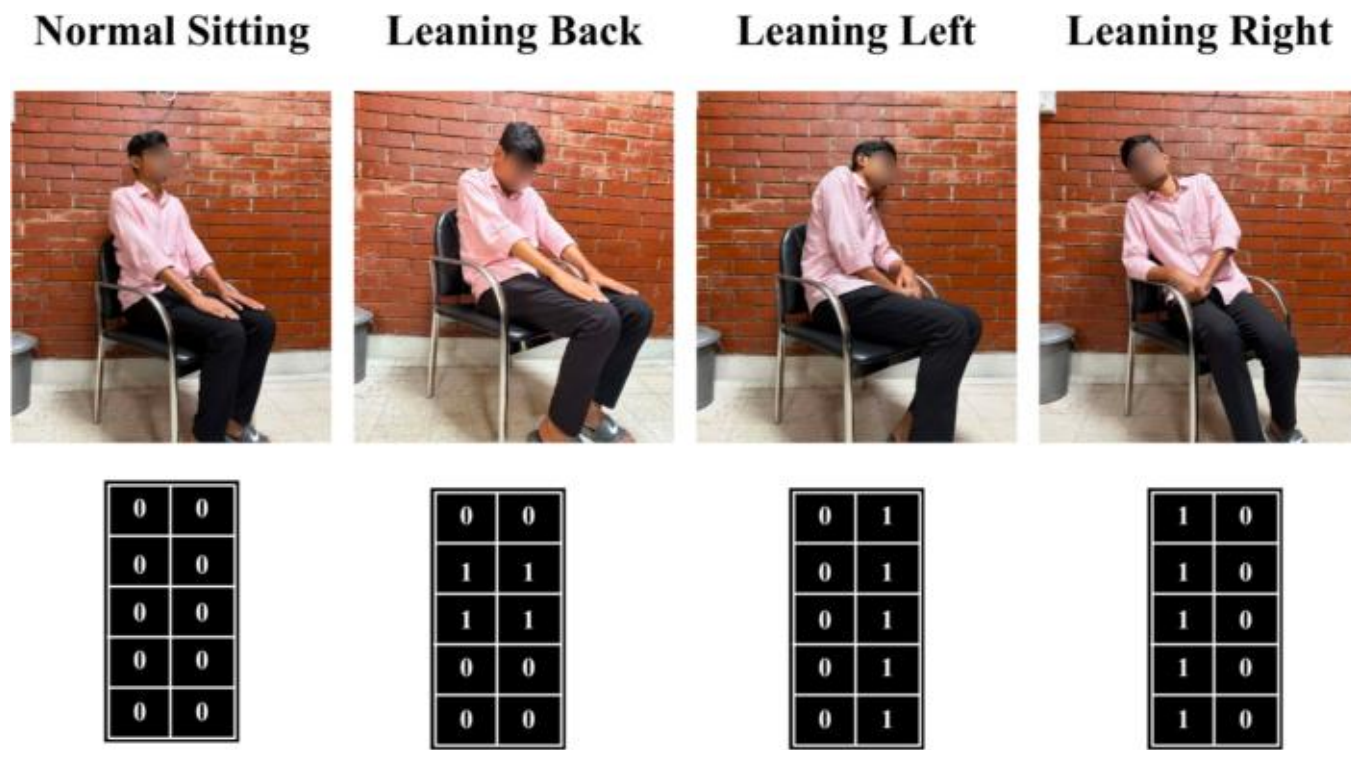

Fig. 3. Representative sitting postures used in the study, along with their corresponding binary switch activation matrices.

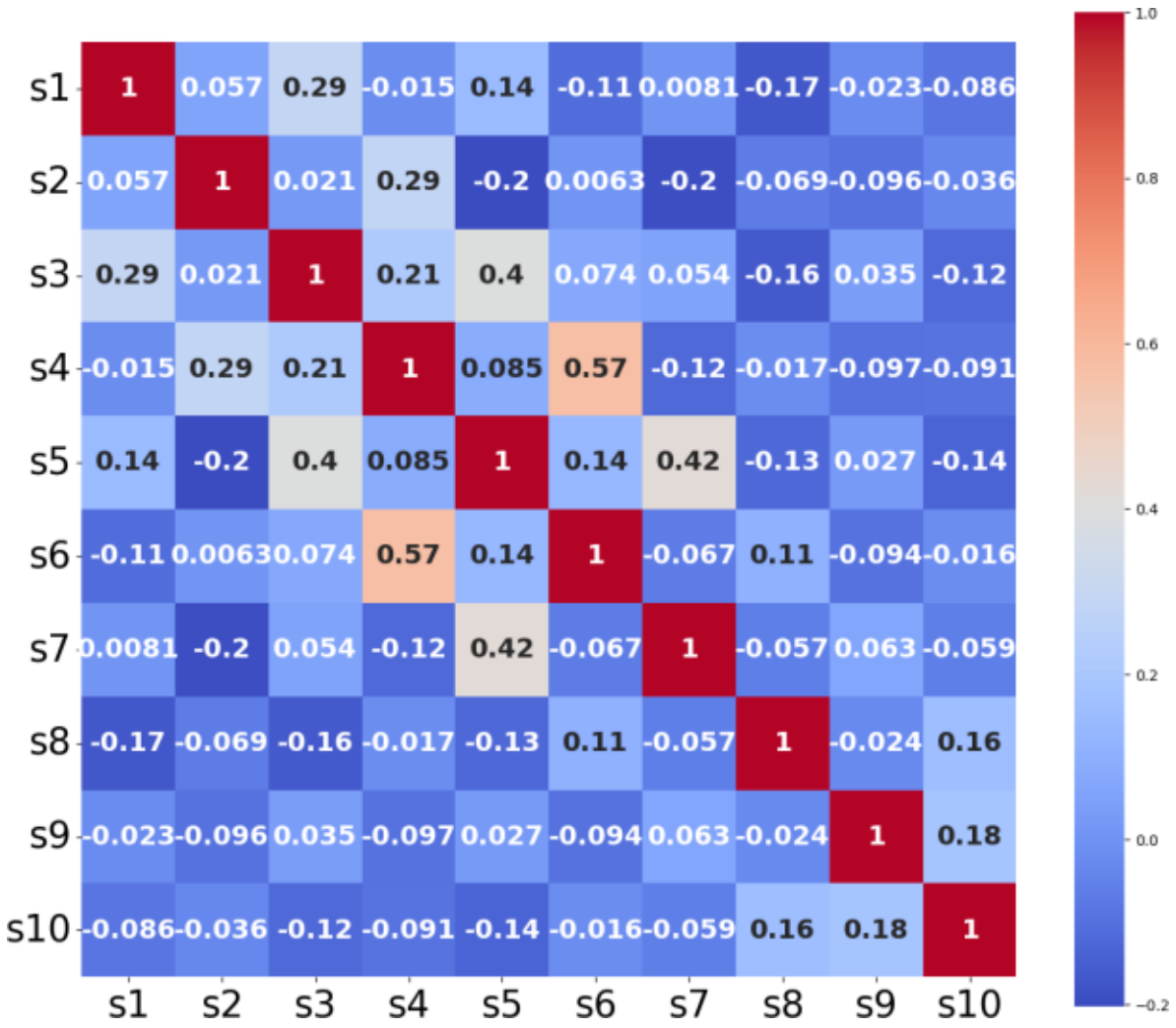

Fig. 4. Correlation matrix of the switch positions

The correlation matrix in Fig. 4 shows that the odd numbered switches (right column) have very weak correlation with the even numbered switches (left column). This reveals that the switches column are working independently. The degree of redundancy is low, as the switches are showing patterns different from each other. Some switches, which are adjacent to each other, show moderate correlation coefficient values such as S4 and S6 ($r = 0.57$), S5 and S7 ($r = 0.42$), and S3 and S5 ($r = 0.40$). During a posture, the back region can focus on a particular switch, and at the same time there is contact on the adjacent switches which can turn on involuntarily. The moderate correlation can be attributed to this phenomenon. Nevertheless, the correlation values are quite below the unity, confirming that the switches are providing discriminative posture information, avoiding any duplication. Consequently, the results show that the ten switch positions are appropriate in posture recognition, as they are not excessively correlated, not redundant, and can collectively capture different contact regions.

TABLE II
Performance Comparison of Different Machine Learning Algorithms[a]

| Algorithm | Accuracy | Precision | Recall | F1-score |
|---|---|---|---|---|
| Random Forest | 93.95 | 94 | 94 | 94 |
| SVM | 94.56 | 95 | 95 | 95 |
| KNN | 91.53 | 94 | 94 | 94 |
| XGBoost | 95.59 | 96 | 96 | 96 |
| Decision Tree | 96.08 | 96 | 96 | 96 |
| Logistic Regression | 96.08 | 96 | 96 | 96 |

[a]All reported values are expressed as percentages (%).

Prior to classifier training, feature normalization was applied to algorithms sensitive to feature scaling. Specifically, the input features for SVM, KNN, and LR were standardized using Python's StandardScaler. It converts each feature to have a mean of 0 and a standard deviation of 1. RF and LR algorithms use a distance-based approach; if the magnitude of some features is greater than the others, then this can create discrepancies in the prediction model. Feature scaling helps to mitigate this. On the other hand, the other classifiers such as DT, SVM, and KNN use a tree-based approach, considering a threshold instead of a distance. As the dataset is already binary, feature scaling is not necessary for these algorithms.

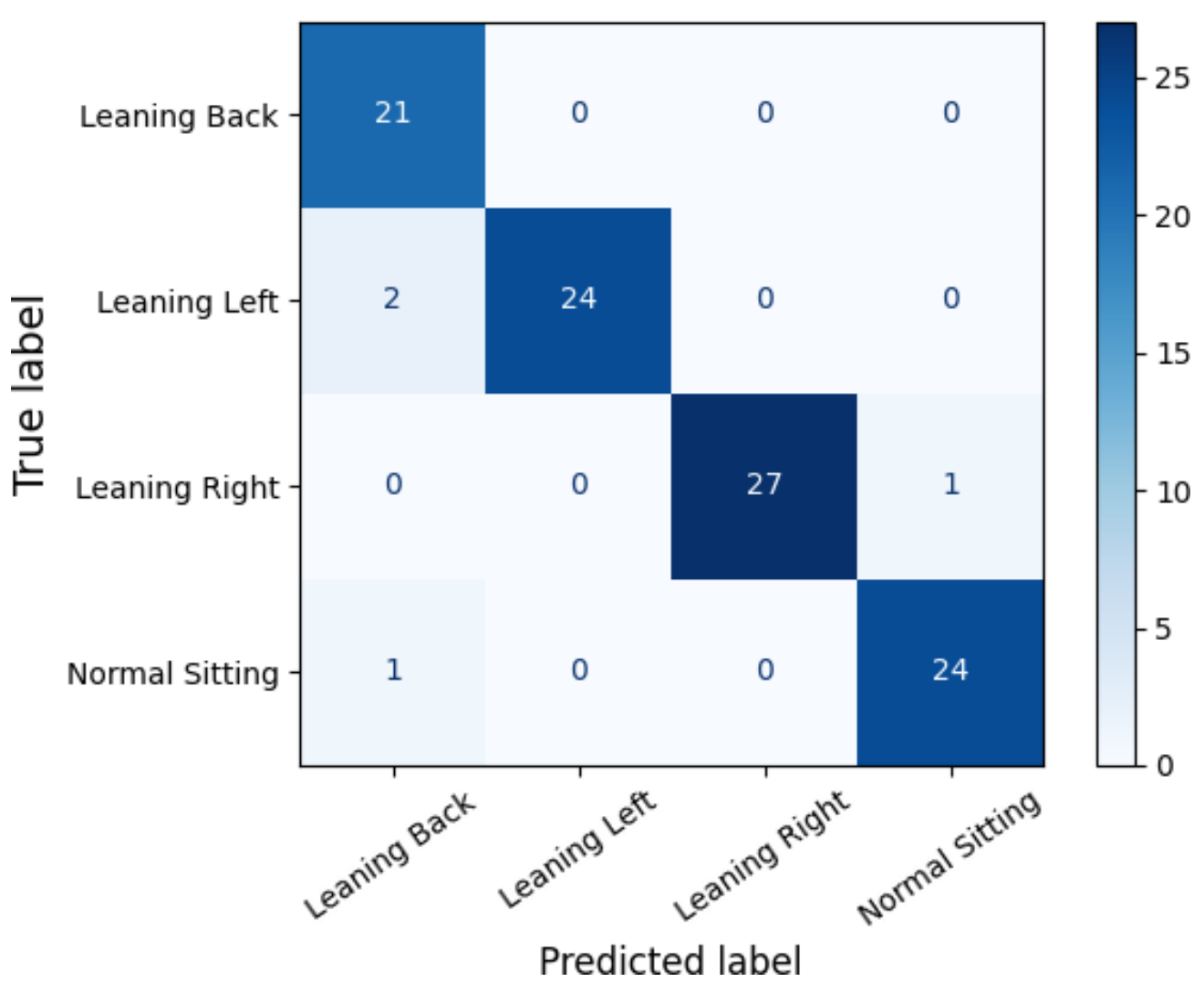

Fig. 5. Confusion matrix for the four classes

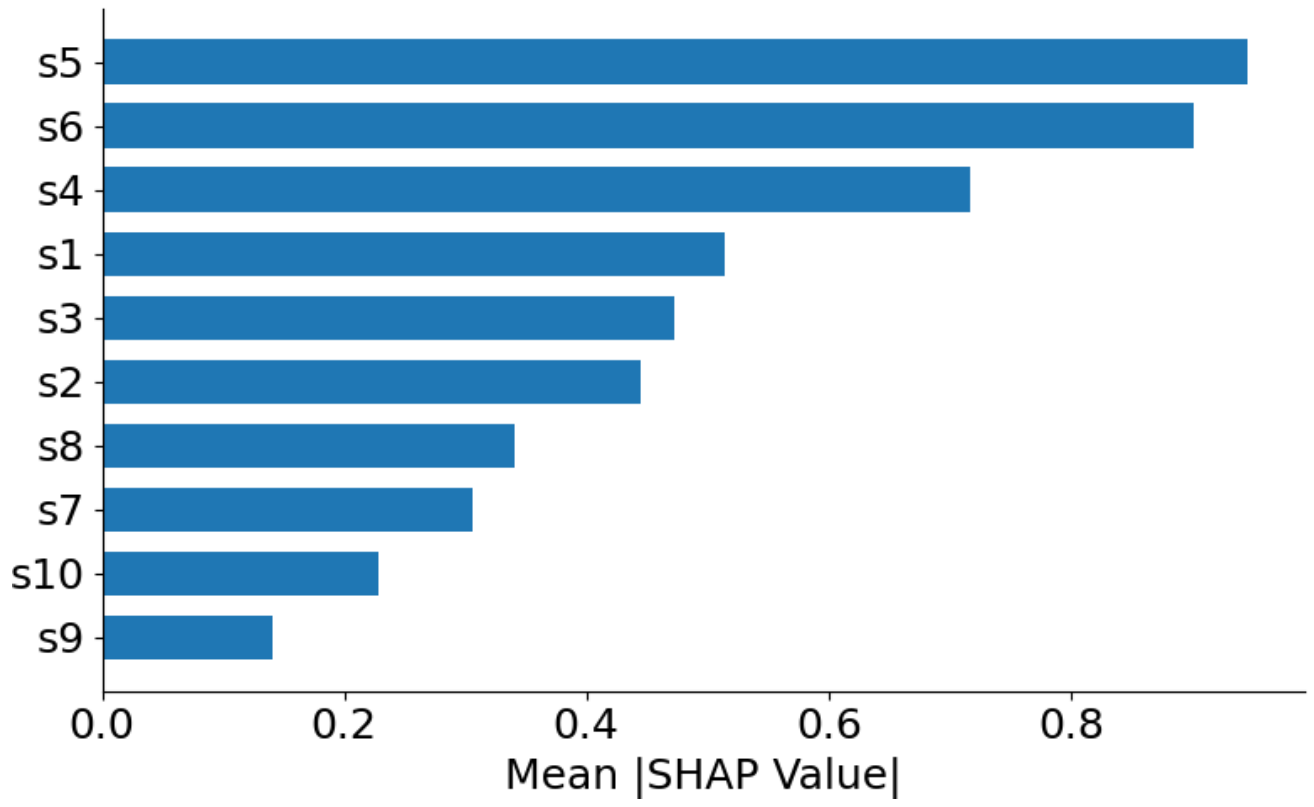

Fig. 6. SHAP plot of switch positions importance

The dataset was randomly divided into 80% training and 20% testing, resulting in 99 training samples and 25 testing samples. 10-fold stratified cross-validation was applied to the dataset, where it was divided into 10 segments, of which 9 segments were trained, and the remaining one segment was tested, keeping the original class distribution equal. This was carried out repeatedly until all the segments had been trained and tested alternatively. This removes the bias that can appear with normal splitting. The stratified approach was chosen over normal K-fold validation to ensure the each divided segment has balanced samples across the 4-classes.

The quantitative performance of the evaluated classifiers is summarized in Table II. All the algorithms achieved accuracies above 91%, showing that all the the models accurately identifies the 4 postures. It verifies that the 10-switch arrangement is compatible with the experiment. The precision, recall, and F1 scores are consistently high, showing accurate class separability and minimal false positives and false negatives. Among them, LR and DT achieved the highest accuracy of 96.08%. DT and LR are simple classifiers, and the other classifiers are more complex; both type of classifiers yield high accuracy, showing the robustness of the system.

Fig. 5 illustrates the confusion matrix obtained from the best-performing classifier. Overall, the confusion matrix exhibits strong diagonal dominance, confirming that the proposed system correctly classifies the majority of posture samples. One out of 26 samples from the Leaning left position was misclassified as Leaning back; this indicates that the subjects leaned more towards the left while leaning on their backs. And 1 out of 26 samples of normal sitting was misclassified with leaning back, indicating subjects leaning back while trying to sit normally.

The Fig. 6 reveals that the switches s1, s4, s5, and s6 are the most important in decision-making. These switches are in the middle of the switchboard, and almost all the postures causes the participants to press these switches. The switches at the top are of secondary importance, and the switches at the lower positions shows the least contribution in determining a posture. This analysis reveals that there is scope to optimize the switch positions of switches S7 to S10.

Fig. 7 illustrates the switch activation frequency for each class. The dark color shows higher frequency. In leaning left, even-numbered switches are activated more (30.7 - 56.7%) compared to the odd-numbered switches (0 - 10.2%). The opposite pattern is observed in the leaning right, where the odd-numbered switches are activated more (36.9 - 70.9%) compared to the even-numbered switches (0 – 9.2%). These positions require subjects to bend in any one direction, and that's why switches from only one side are being activated. For the leaning back position, the highest activation happens for switches S3 to S6, with S6 being the highest (80%).This behavior confirms that when the subjects lean back, only the middle back is in contact with the sensor board. The Normal position has the lowest activation, as there is not much pressure exerted on the switches, and they do not get turned on. In general, the middle switches have higher activation than the upper and lower switches. Switch S5 has almost similar activation for leaning back (68.6%) and right (70.9%), showing that people lean towards right more when they lean back.

Overall, the experimental results demonstrate that the proposed contact sensor-based posture recognition system provides reliable and accurate classification while maintaining a simple hardware architecture. Unlike conventional pressure-sensing systems, which require force-sensitive elements, capacitive structures, or piezoresistive materials that may exhibit calibration drift, nonlinear responses, material degradation, and increased manufacturing cost, the proposed system utilizes low-cost binary contact switches with high mechanical durability and minimal signal conditioning requirements. Since the objective of this work is to identify postures rather than pressure distribution, binary contact information is sufficient to achieve high recognition accuracy. While pressure sensor arrays provide detailed pressure maps and center-of-pressure information that are advantageous for biomechanical analysis and clinical gait or seating studies, our study demonstrates that such high-dimensional information is not required for coarse posture recognition. The proposed work therefore offers a simple posture recognition system with minimal implementation cost and high accuracy, making it suitable for long-term posture monitoring applications.

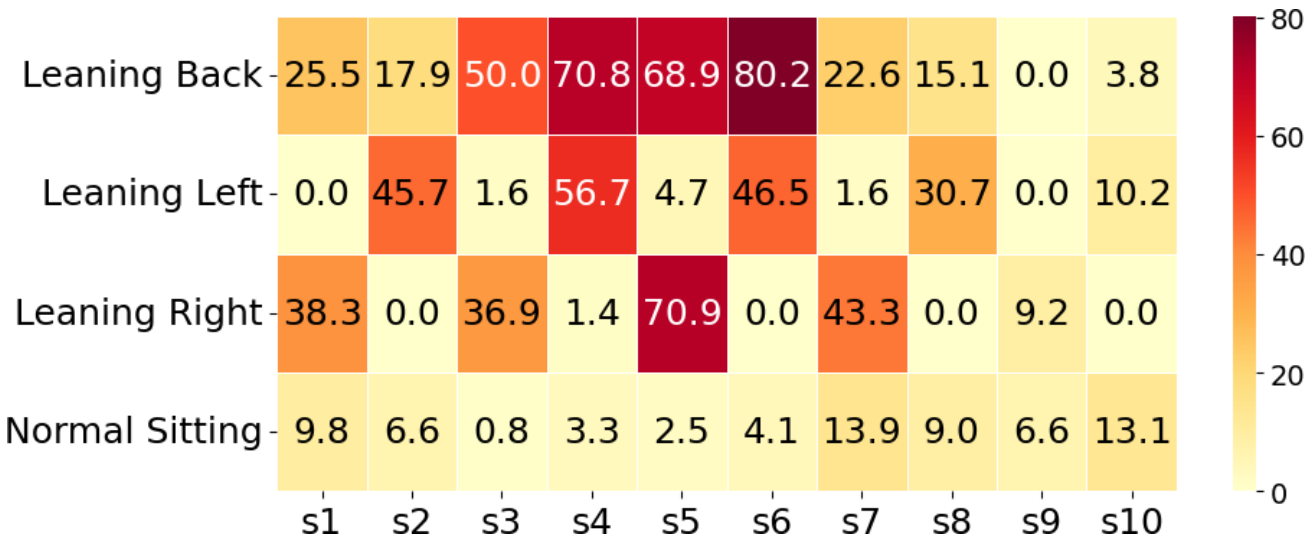

Fig. 7. Activation frequency matrix of switch position for each posture class

## V. Conclusion

This work demonstrates that reliable sitting posture recognition does not necessarily require pressure measurement. Binary contact switches can capture the spatial distribution of body-chair interaction with sufficient information to distinguish between common sitting postures. This shift from pressure magnitude sensing to contact pattern sensing offers an alternative design philosophy for posture monitoring. The mechanical switch-to-binary based approach simplifies both hardware and signal processing by excluding analog signal conditioning and calibration. Future work will investigate a wider range of dynamic postures, recording history of body postures through IoT systems, and real-time feedback for posture correction. The proposed contact sensing framework provides a practical foundation for smart chairs for rehabilitation and home healthcare.